# Structural Transition of Actin Filament in a Cell-Sized Water Droplet with a Phospholipid Membrane

M. Hase and K. Yoshikawa[a]

*Department of Physics, Graduate School of Science, Kyoto University, Kyoto 606-8502, Japan*

[a] *Electronic mail: yoshikaw@scphys.kyoto-u.ac.jp*

**Abstract**

Actin filament, F-actin, is a semiflexible polymer with a negative charge, and is one of the main constituents on cell membranes. To clarify the effect of cross-talk between a phospholipid membrane and actin filaments in cells, we conducted microscopic observations on the structural changes in actin filaments in a cell-sized (several tens of micrometers in diameter) water droplet coated with a phospholipid membrane such as phosphatidylserine (PS; negatively-charged head group) or phosphatidylethanolamine (PE; neutral head group) as a simple model of a living cell membrane. With PS, actin filaments are distributed uniformly in the water phase without adsorption onto the membrane surface between 2 and 6 mM $Mg^{2+}$, while between 6 and 12 mM $Mg^{2+}$, actin filaments are adsorbed onto the inner membrane surface. With PE, actin filaments are uniformly adsorbed onto the inner membrane surface between 2 and 12 mM $Mg^{2+}$. With both PS and PE membranes, at $Mg^{2+}$ concentrations higher than 12 mM, thick bundles are formed in the bulk water droplet accompanied by the dissolution of actin filaments from the membrane surface. The attraction between actin filaments and membrane is attributable to an increase in the translational entropy of counterions accompanied by the adsorption of actin filaments onto the membrane surface. These results suggest that a microscopic water droplet coated with phospholipid can serve as an easy-to-handle model of cell membranes.

**INTRODUCTION**

Actin filament is a constituent of the cytoskeleton and plays an important role in the cell morphology, mechanical stability, and motility.[1,2] Actin filament shows a variety of morphologies in cytoplasm, especially in conjunction with membrane. It has been considered that actin filaments form assembly together with structural proteins such as α-actinin, fimbrin and scruin.[3-7] It is also known that actin filaments attach to the surface of the membrane and form a meshwork structure through the mediation of structural proteins such as CD44 and ezrin-radixin-moesin (ERM) proteins.[8,9]

Recently, it has been reported that actin filaments, even in the absence of such structural proteins, form bundles in the presence of multivalent cations due to the nature of actin filament as a semiflexible polyelectrolyte.[10,11] In addition, it has been shown that, in a giant liposome made of phosphatidylcholine (PC; neutral head group), actin filament is adsorbed onto membrane in the presence of $Mg^{2+}$ without any structural protein.[12,13]

In the plasma membranes of eukaryotic cells, phospholipids are distributed asymmetrically between the inner and outer layers.[14,15] It is expected that this asymmetry is important for cell function.[16-19] In the red cell membrane, PC is mostly found in the outer membrane, while phosphatidylserine (PS) and phosphatidylethanolamine (PE) are found in the inner membrane.[14] In addition, negatively-charged lipids such as PS and phosphatidylinositol (PI) are mostly found in the inner membrane of yeast cells.[15] Thus it is expected that actin filaments receive strong electrostatic repulsion from the negatively-charged inner membrane. It has been suggested that dynamic change in lipid distribution in the inner membrane play a role in the adsorption and desorption of actin filaments on the inner membrane surface during cell division.[19] However, the mechanisms that control this adsorption and desorption of actin filaments to the inner surface of the cell membrane are poorly understood.

Several studies have been conducted on giant liposomes encapsulating actin filaments,[12,13,20-23] and giant liposomes made of mainly PC have been most frequently used as a model cell. Currently, it is known that PC is suitable as the major component for preparing giant liposomes. However, the method for preparing giant liposomes with a composition resembling that of the inner cell membrane has not yet been established.

In this study, we used a cell-sized water droplet with a lipid membrane (CWD). We consider the water phase of CWD as a model of the cytoplasm because the head of the lipid at the water-oil interface faces the water phase. We conducted a microscopic observation of CWD encapsulating actin filaments in the presence of $Mg^{2+}$ to better understand the interaction between actin filaments and a model membrane composed of PS and PE.

**EXPERIMENTAL SECTION**

Monomeric actin (G-actin, molecular weight (MW) of 42kDa) was prepared from rabbit skeletal muscles by the method of Spudich and Watt[24]. Purified actin was stored in G-buffer (2 mM Tris/HCl, pH 8.0, 0.2 mM $CaCl_2$, 0.2 mM ATP, 0.5 mM mercaptoethanol). In the presence of 2mM $Mg^{2+}$, G-actin polymerizes into filament, F-actin, within several minutes. Actin filament was stained by rhodamine phalloidin and observed by a confocal laser-scanning microscope (LSM 510, Carl Zeiss) with a He-Ne laser. Excitation was performed at $\lambda = 543$ nm and observations were made at $\lambda > 585$ nm with a slit width of 7.3 μm.

The lipids, 1,2-dioleoyl-sn-glycero-3-phatidylserine (DOPS) and

1,2-dioleoyl-sn-glycero-3- phosphatidylethanolamine (DOPE), were dissolved in mineral oil and sonicated for 1 hour. To prepare CWD with a lipid membrane, 1 μl F-buffer (2 mM Tris/HCl, pH 8.0, 0.2 mM $CaCl_2$, 4-40 mM $MgCl_2$, 0.2 mM ATP, 0.5 mM mercaptoethanol) was added to 200 μl mineral oil in the presence of 1mM lipid, and 1 μl G-actin was then added to this droplet and quickly mixed. CWD with a diameter of several tens of μm was obtained after several seconds of vortex agitation (Fig.1). Actin filaments formed spontaneously in CWD after it had been allowed to stand for several minutes.

Results

(i) Structural changes in actin filaments due to $Mg^{2+}$ in bulk solution

Figure 2 shows typical confocal images of actin filaments labeled by rhodamine phalloidin at 2 mM and 20 mM $Mg^{2+}$. At 2 mM $Mg^{2+}$, a single actin filament exhibited thermally agitated undulation. At 20 mM $Mg^{2+}$, thick bundles of actin filaments were formed at such a high actin density that the actin bundles overlapped and stuck to each other to form a rigid network.

Figure 3 shows light-scattering plot made with a spectrometer to monitor actin filament bundle formation by $Mg^{2+}$. The incident wavelength was set to 450 nm and the scattered light was detected at 450 nm. Accompanied by the formation of bundles, the scattering intensity showed a steep increase in the vicinity of 12 mM $Mg^{2+}$. Such a change in the scattering intensity corresponds to the experimental trend reported previously.[11]

(ⅱ) **Structural changes in actin filaments by $Mg^{2+}$ in CWD with a PS membrane**

Figure 4 shows typical confocal images of actin in CWD with a PS membrane at several $Mg^{2+}$ concentrations. In the absence of $Mg^{2+}$, G-actins were uniformly distributed in CWD. At 2 mM $Mg^{2+}$, actin filaments formed and were uniformly distributed in the water phase without adsorption on the membrane surface. As the $Mg^{2+}$ concentration increased, actin filaments tended to adsorb onto the membrane at $Mg^{2+}$ concentrations below 12 mM and accumulated on the membrane surface. At $Mg^{2+}$ concentrations higher than 12 mM, bundles of actin filaments were formed in CWD, and the number of actin filaments attached to the membrane decreased. At a high actin concentration, bundles stuck to each other and a network of actin filaments was formed in CWD.

(ⅲ) **Structural changes in actin filaments by $Mg^{2+}$ in CWD with a PE membrane**

Figure 5 shows typical confocal images of actin in CWD with a PE membrane at several $Mg^{2+}$ concentrations. In the absence of $Mg^{2+}$, G-actins were uniformly distributed in the water phase of CWD. At 2 mM $Mg^{2+}$, actin filaments formed and were adsorbed onto the PE membrane. According to a previous report, PC membrane shows a similar tendency and it has been suggested that PC membrane has a positive charge in the presence of $Mg^{2+}$ and this is accompanied by the adsorption of actin filaments on the membrane.[12,25] At $Mg^{2+}$ concentrations higher than 12 mM, bundles of actin filaments were formed in CWD, and the number of actin filaments attached to the membrane decreased, as with a PS membrane,.

Figure 6 shows diagram of the structural changes in actin filaments at different $Mg^{2+}$ concentrations in the bulk solution or in the water phase of CWD with PS and PE membranes.

**DISCUSSION**

(ⅰ) **Structural changes in actin filaments by $Mg^{2+}$ in CWD with a negatively-charged membrane**

Actin filaments are negatively-charged and exhibit strong electrostatic repulsion with a negatively-charged membrane. In the past, various experiments have been conducted in giant liposomes to examine the interaction between actin filaments and membrane.[12,13,20-23] However, there have been few studies on the interaction between actin filaments and negatively-charged membrane in giant liposomes because of the difficulty of preparing giant liposomes composed of a single type of negatively-charged lipid. In this study we examined the interaction between actin filaments and negatively-charged membrane through the use of CWD with a single species of PS membrane, where the head of PS, with a negative charge, faces the water phase in CWD.

Actin filaments are known to self-assemble into bundles, which consist of parallel densely packed rods. Actin filaments are negatively charged and their association is induced by divalent cations such as $Mg^{2+}$.[10,11] The Poisson-Boltzmann theory is frequently used to understand the characteristics of polyelectrolytes.[26,27] However, this mean-field approximation fails to explain the formation of stable parallel contact-alignment of the polyelectrolytes, since this theory predicts that like-charge polyelectrolytes will always exhibit repulsive interaction. Recently, studies on the correlation of counterions have shown the stabilization of parallel contact-alignment between like-charged polyelectrolytes.[28–32] Some of these theoretical studies have adopted the term "attraction between like-charge". In contrast to this term, a stiff polyelectrolyte must undergo a discrete transition between a dispersed disordered state and a condensed ordered state;[33] i. e., fundamentally, the assembly process exhibits the general features of a first-order phase transition under the framework of Landau's symmetry argument[34]. This indicates that the notion of repulsive and attractive interaction between like charges is inaccurate. At least from a physical perspective, this interaction should be interpreted in terms of the change in free energy. For example, the bimodality of free energy in the assembly process means that the pair-wise interaction between like charges remains repulsive even at the transition point.

Thus, in the present paper we will consider the structural changes in actin filaments and membrane within the framework of free energy. For the free energy before and after the structural transition, we simply regard the contributions of the translational entropy of the counterions and electrostatic energy. The free energy for a filament would be written as

$$F = F_{tra} + F_{ele} \tag{1}$$

where $F_{tra}(=-TS_{tra})$ corresponds to the translational entropic contribution and $F_{ele}(=U_{ele})$ is the electrostatic energy.

(ⅱ) **Adsorption of actin filaments onto a PS membrane**

We consider that an actin filament is adsorbed onto a PS membrane. We consider that the change in translational entropy of actin filament is quit small compared to that of counterions. The translational entropy of freely moving small ions (bivalent cations,

monovalent cations and monovalent anions) in a solution of actin filament can be written as[35]

$$S_{tra}/k_B = -\sum_{i=2+,1+,1-}(P_i^{in}\ln(P_i^{in}/V^{in})) - \sum_{i=2+,1+,1-}(P_i^{out}\ln(P_i^{out}/V^{out})) \quad (2)$$

where $V^{in}$ and $V^{out}$ are the effective volumes of ions fluctuating inside and outside, respectively, the actin filament and membrane; in other word, condensed and decondensed ions.[36] The contribution of small negative ions to $\Delta S_{tra}$ would be negligible. Next, we assume that the change in translational entropy is due to ion exchange between monovalent cation and the bivalent cation $Mg^{2+}$. $\Delta S_{tra}$ would be given as

$$\Delta S_{tra}/k_B = -P_{2+}\ln(V^{out}/V^{adsorption}) + P_{1+}\ln(V^{out}/V^{membrane+filament}) \quad (3)$$

where $P_{2+}$ and $P_{1+}$ are the number of $Mg^{2+}$ and monovalent ions that can be exchanged with the adsorption transition ($P_{1+} = 2P_{2+} = \alpha$). $V^{adsorption}$ and $V^{filament+membrane}$ are the effective volumes of ions fluctuating in the states in which an actin filament is adsorbed onto and desorbed apart from the membrane, respectively, and would be written as

$$\begin{aligned} V^{adsorption} &= V^{filament+membrane} - V^{overlap} \\ V^{filament+membrane} &= \pi(r+\lambda_D)^2 L + 4\pi a^2 \lambda_D / N_{total} \end{aligned} \quad (4)$$

where $V^{overlap}$ is the overlapping part of the effective volume of actin filament and the membrane, $a$ is the radius of CWD, $r$ and $L$ are respectively the radius and length of actin filament and $N_{total}$ is the total number of actin filaments in CWD. We take the

Debye length $\lambda_D$ for the depth of the ionic environment for $V^{filament+membrane}$. We adopt the data previously determined for actin filament; $L = 14[nm]$, $r = 4[nm]$, actin filament consists of 370 monomers per μm unit length, and the linear charge density is $4e/nm$ [37,38], where we consider that the difference in the length of each actin filament is not important for this discussion. In the case of 10 mM $Mg^{2+}$, 350 μg / ml actin and $a = 40 \mu m$, $\alpha \sim 10^5$ ( the number of negative moieties of actin filament and membrane), from (8) and (9) $\Delta S_{tra}$ is calculated to be

$$\Delta S_{tra}/k_B \sim 3 \times 10^5 \tag{5}$$

When an actin filament is absorbed onto the membrane, the change in translational entropy upon the exchange of counterions is a large positive value and contributes to the attraction between actin filament and the membrane.

The electrostatic energy of actin filament and membrane can be written as

$$U_{ele} = \frac{\rho_f^2 r^4 L \pi}{4\varepsilon_0}(\frac{1}{4\varepsilon_1} + \frac{1}{\varepsilon_2}\ln\frac{r+\lambda_D}{r}) + \frac{\pi a^2 \rho_m^2 \lambda_D}{2\varepsilon_0 N_{total}}(\frac{1}{\varepsilon_2} + \frac{1}{\varepsilon_3}) \tag{6}$$

where the first term is the electrostatic energy of actin filament and the second term is that of the membrane, $\rho_f$ and $\rho_m$ are the remaining charges in the actin filament and membrane, respectively,, and $\varepsilon_0$, $\varepsilon_1$, $\varepsilon_2$ and $\varepsilon_3$ are the dielectric constants of vacuum, actin filament, water and oil, respectively. We consider that $\rho_m$ is sufficiently small to neglect the electrostatic energy of the membrane and that $\varepsilon_1 \approx 4, \varepsilon_2 = 80$, and

$\varepsilon_3 \approx 2$.[39] The change in electrostatic energy before and after adsorption is calculated to be

$$\Delta U_{ele}/k_B T \sim 1 \times 10^3 \tag{7}$$

From (5) and (7), the change in free energy before and after adsorption ($\Delta F \approx \Delta U_{ele} - T\Delta S_{tra}$) is a negative value, and adsorption to the membrane occurs.

### (ⅲ) Formation of bundles of actin filaments in CWD

We consider that N actin filaments form a bundle. As in adsorption to the membrane, the change in $\Delta S_{tra}$ would be given as

$$\Delta S_{tra}/k_B = -P_{2+} \ln(V^{out}/V^{bundle}) + P_{1+} \ln(V^{out}/V^{filament}) \tag{8}$$

where $V^{bundle}$ and $V^{filament}$ are the effective volumes of the bundle state and actin filament state, respectively, and can be written as

$$\begin{aligned} V^{bundle} &= \pi(R+\lambda_D)^2 L/N \\ V^{fialment} &= \pi(r+\lambda_D)^2 L \end{aligned} \tag{9}$$

where $R$ is the radius of a bundle ($R \approx r\sqrt{N}$). Next, the change in electrostatic energy $\Delta U_{ele}$ would be written as

$$\Delta U_{ele} = U_{ele}^{bundle} - U_{ele}^{filament} \tag{10}$$

$$\begin{aligned} U_{ele}^{bundle} &= \frac{\rho_b^2 R^4 L\pi}{4\varepsilon_0}(\frac{1}{4\varepsilon_1} + \frac{1}{\varepsilon_2}\ln\frac{R+\lambda_D}{R})\Big/N \\ U_{ele}^{filament} &= \frac{\rho_f^2 r^4 L\pi}{4\varepsilon_0}(\frac{1}{4\varepsilon_1} + \frac{1}{\varepsilon_2}\ln\frac{r+\lambda_D}{r}) \end{aligned} \tag{11}$$

where $\rho_b$ is the remaining charge of a bundle. From (8) ~ (11), optimization of the change in free energy ($\Delta F \approx \Delta U_{ele} - T\Delta S_{tra}$) leads to a finite size of the bundle

$$R \approx \sqrt{\frac{\rho_f^2 r^2 L\pi/8\varepsilon_0\varepsilon_1 + 2k_B TP_{1+}\ln(\sqrt{V^{bundle}V^{out}}/V^{filament})/r^2}{\rho_b^2 L\pi/4\varepsilon_0\varepsilon_1}} \quad (12)$$

where we consider that $\varepsilon_1 \ll \varepsilon_2$ and $R \gg r (=4[nm]) \sim \lambda_D (=1.24[nm]$, in the case of 20 mM $Mg^{2+}$). We consider that $\alpha \approx 5.6\times10^4$ ( the number of negative moieties of actin filament). From (8) ~ (12), the change in free energy is calculated to be

$$\Delta F/k_B T \approx -2\times 10^5 \quad (13)$$

Thus, actin filaments undergo a bundle transition in CWD. It is expected that it is difficult for a rigid bundle to attach to a membrane with high curvature due to the penalty of bending energy ( $= \kappa \frac{NL}{a^2} = L_P \frac{NL}{a^2} k_B T \approx 0.15 N k_B T \gg k_B T$, in the case of $R \gg r$), where $\kappa$ and $L_P (=17[\mu m])$ are the bending modulus and the persistence length of an actin filament, respectively.[38]

Each actin filament that comprises a bundle differs in length, and the bundle still has parts that actin filaments can adsorb to. Thus, bundles can form network structures at high actin density to the point where bundles overlap.

(ⅳ) **Adsorption of actin filaments to PE membrane**

Actin filaments attach to PE (neutral lipid) membrane in the presence of 2 mM

$Mg^{2+}$. As previously reported, a PC (neutral lipid) membrane of a giant liposome shows a similarly tendency,[12,13] where PC exhibits a positive charge in the presence of divalent cations such as $Mg^{2+}$ and $Ca^{2+}$.[25,40] The head of PE has an electric dipole similar to the head of PC. PE is expected to exhibit a positive charge in the presence of divalent cations such as $Mg^{2+}$ and $Ca^{2+}$, which would cause attractive interaction between the membrane surface and actin filaments.

**CONCLUSION**

We considered a use of cell-size water droplet with a lipid membrane (CWD) as a model cell. We conducted confocal microscopy observations of CWD with a lipid membrane encapsulating actin filaments to investigate the interaction between actin filaments and PS or PE membranes at various salt concentrations.

We found that actin filaments can attach to a PE or PS membrane without any structural protein. Furthermore, we could control the adsorption-desorption of actin filaments to the membrane by the sort of membrane (PE and PS) and the salt concentration. Based on a calculation of the free energy of an actin filament and membrane system, we found that this assembly of actin filaments and membrane occurs not only due to the specific nature of protein but also to the nature of actin filaments as charged rigid polymers.

The water phase of CWD used in this paper may be useful as a simple model cellular system compared to giant liposomes for several reasons. First, it is easy to prepare cell-sized CWD with a single species of lipid with a large spontaneous curvature such as PE or PS, which account for a large part of the inner plasma membrane. Second, CWD is very stable against out-field effects, such as osmotic pressure. Third, the water phase of CWD resembles the cytoplasm environment because the head of the lipid at the water-oil interface faces the water phase, and thus a protein such as actin would not denature.

**ACKNOWLEGMENT**

This work was supported by the Grant-in-Aid for the 21st Century COE "Center for Diversity and Universality in Physics" and Grant-in-Aid for Scientific Research on Priority Areas (No.17076007) "System Cell Engineering by Multi-scale Manipulation" from the Ministry of Education, Culture, Sports, Science and Technology of Japan. We thank Mr. S. Watanabe, Department of Physics, Graduate School of Science, Kyoto University, for his kind advice on the experimental procedures and analysis.

**Figure caption**

Fig.1: (a) Schematic representation of the experimental procedure. (b) Diagram of CWD (c) Microscopic image of a CWD.

Fig.2: Confocal images of actin filaments and actin bundles in bulk aqueous solution. (a) 2.0 mM $Mg^{2+}$, 8.0 μg / ml actin. (b) 20 mM $Mg^{2+}$, 160 μg / ml actin. (c) 20 mM $Mg^{2+}$, 700 μg / ml actin.

Fig.3: Formation of an actin bundle as monitored by light scattering. Actin concentration is 350μg / ml.

Fig.4: Confocal images of actin filaments in CWD with a PS membrane depending on the $Mg^{2+}$ concentration. (a) 0 mM $Mg^{2+}$, (b) 2 mM $Mg^{2+}$, (c) 6 mM $Mg^{2+}$, (d) 10 mM $Mg^{2+}$, (e) 12mM $Mg^{2+}$, (f) 20 mM $Mg^{2+}$; a)-f) 350 μg / ml actin. g) 20 mM $Mg^{2+}$, 700 μg / ml actin.

Fig.5: Confocal images of actin filaments in CWD with a PE membrane depending on the $Mg^{2+}$ concentration and 350 μg / ml actin. (a) 0 mM $Mg^{2+}$, (b) 2 mM $Mg^{2+}$, (c)12 mM $Mg^{2+}$, (d)20mM $Mg^{2+}$

Fig. 6: Diagram of the structural changes in actin filaments depending on the $Mg^{2+}$ concentration in bulk solution or in the water phase of CWD with a PS or PE membrane. "G-actin" and "filament" refer to the states in which actin monomer and actin filament, respectively, are uniformly distributed uniformly in bulk solution or the water phase of CWD, "surface adsorption" refers to the state in which actin filaments are adsorbed onto the inner membrane surface and accumulates on the membrane surface, and "bundle" indicates the state in which bundles of actin filaments are formed in the water phase of CWD accompanied by a decrease in the number of actin filaments attached to the membrane.

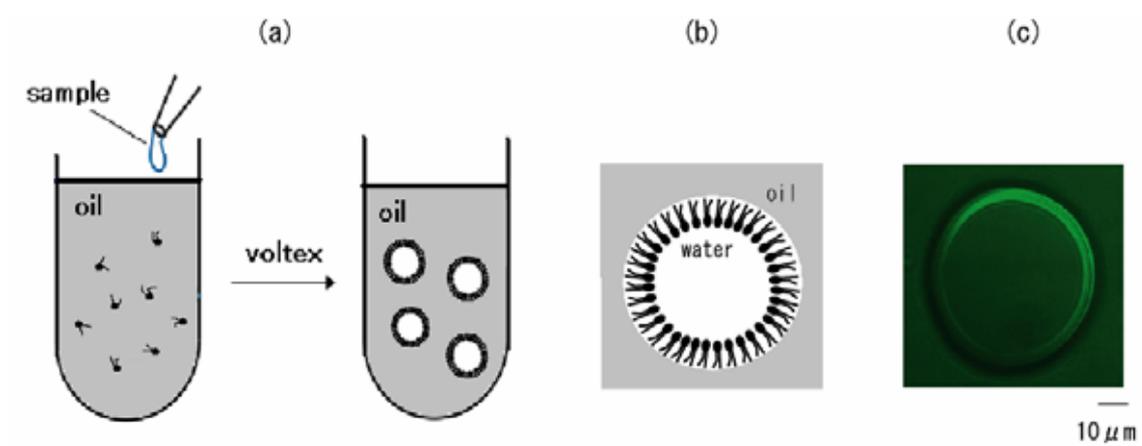

Fig.1

(a) 2 mM Mg$^{2+}$ 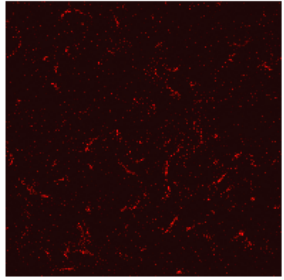 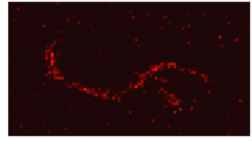

40 μm    10 μm

(b) 20 mM Mg$^{2+}$ 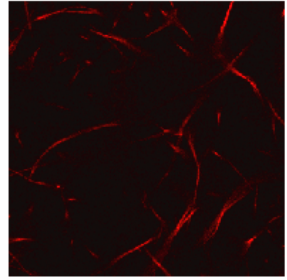 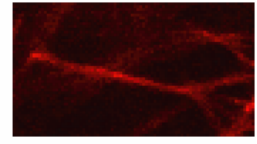

40 μm    10 μm

(c) 20 mM Mg$^{2+}$, high actin concentration 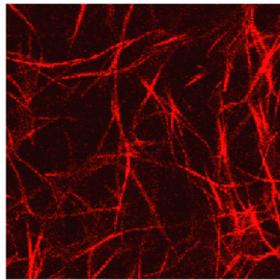

40 μm

Fig.2

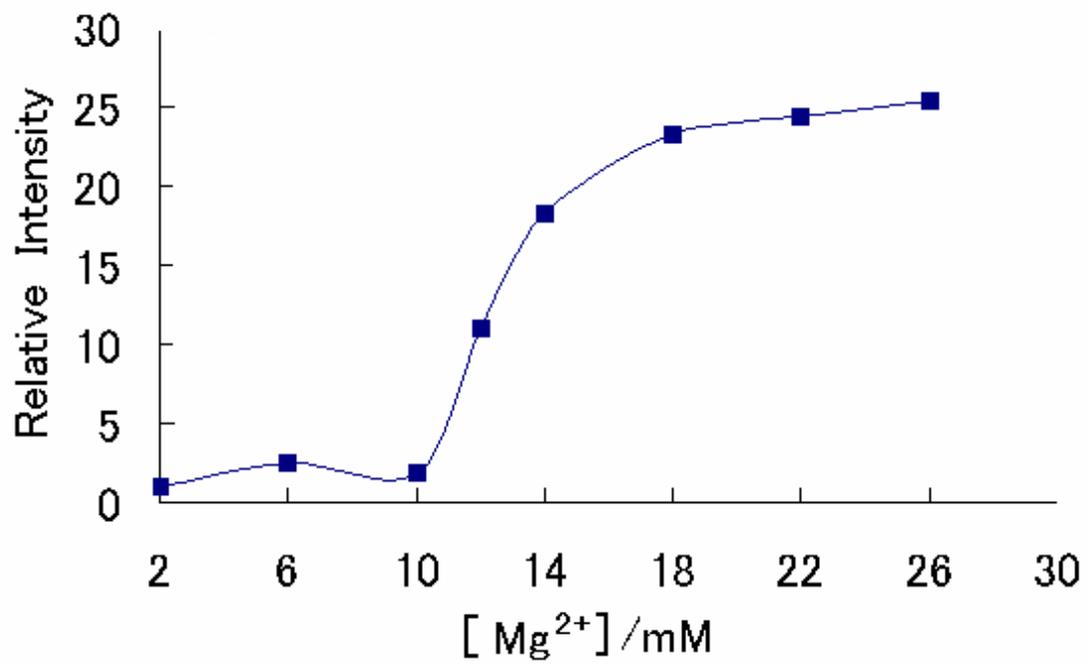
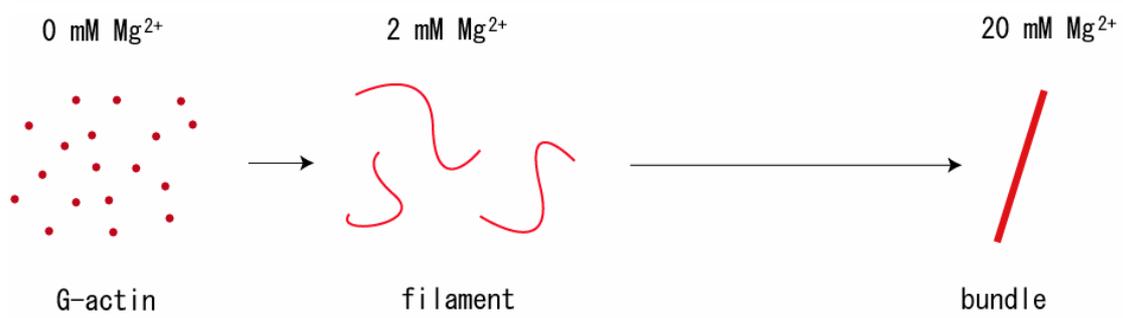

Fig.3

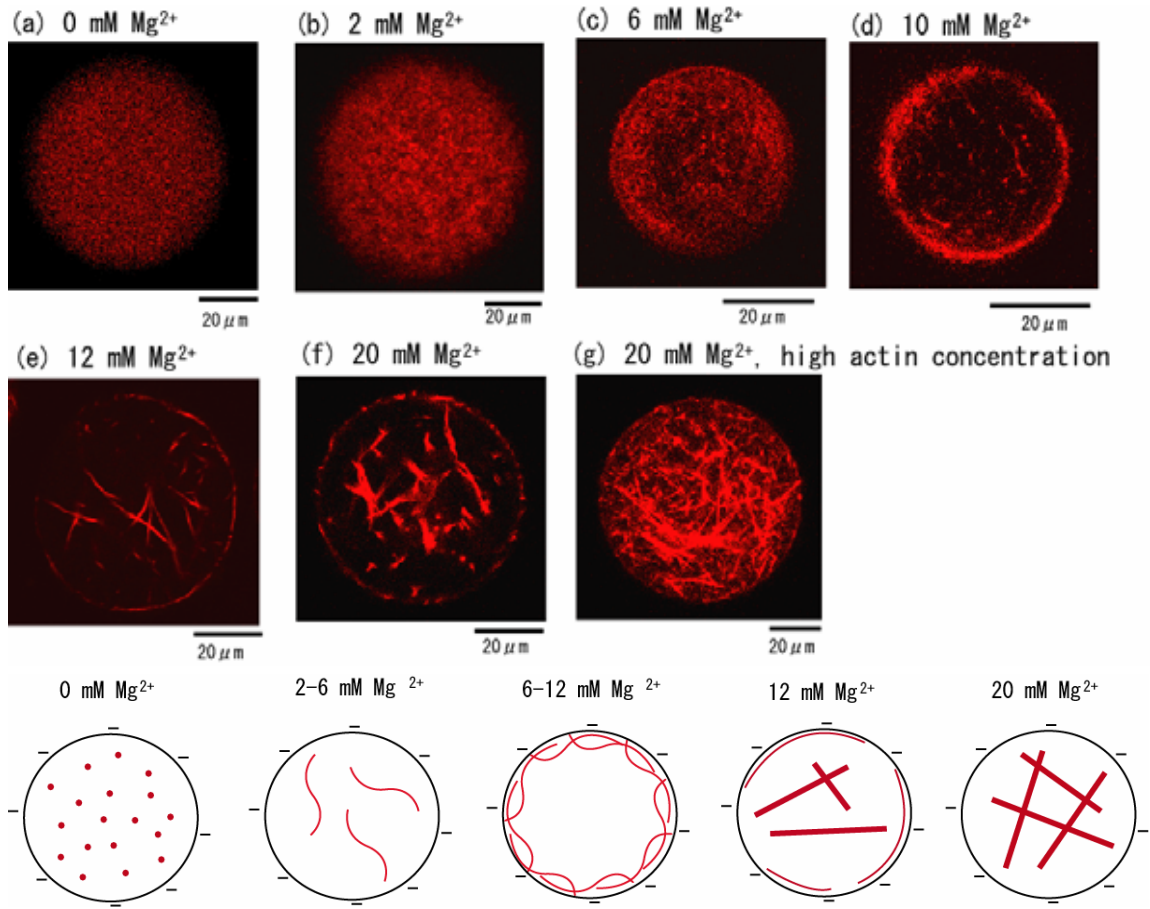

Fig.4

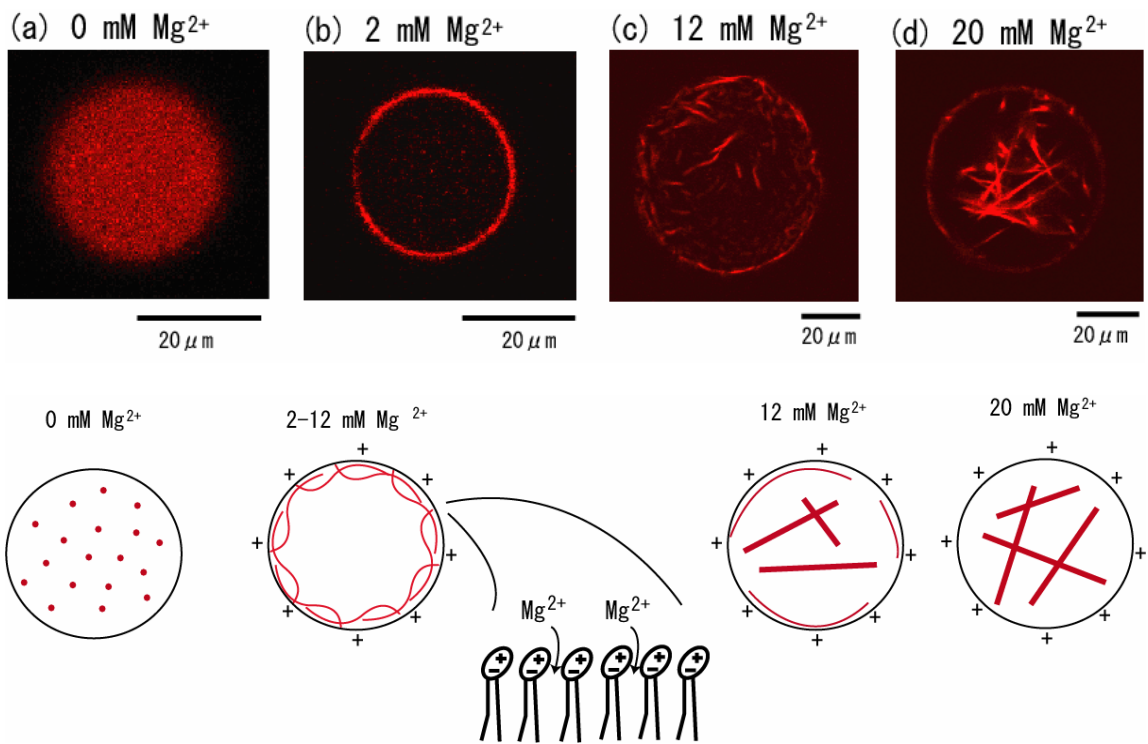

Fig.5

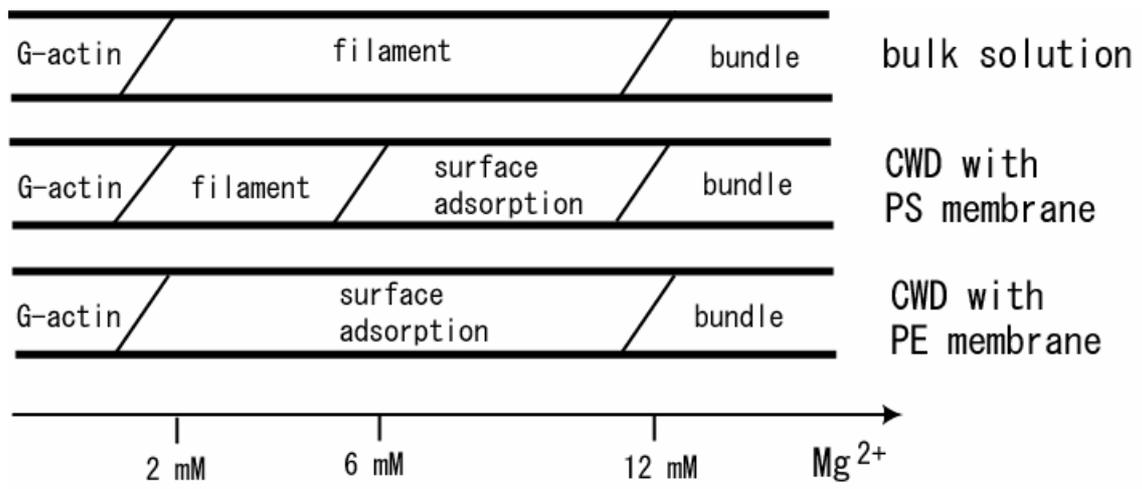

Fig.6